\documentclass{llncs}
\usepackage{graphicx}
\usepackage{subfigure}
\usepackage{url}
\usepackage{common}

\begin{document}
\title{Producing a Unified Graph Representation from Multiple Social Network Views}
\authorrunning{Greene et al.}

\author{Derek Greene \and P\'{a}draig Cunningham}

\institute{School of Computer Science \& Informatics, University College Dublin \\
\email{\{derek.greene,padraig.cunningham\}@ucd.ie} }

\maketitle
%-------------------------------------------------------------

\begin{abstract}
In many social networks, several different link relations will exist between the same set of users. Additionally, attribute or textual information will be associated with those users, such as demographic details or user-generated content. For many data analysis tasks, such as community finding and data visualisation, the provision of multiple heterogeneous types of user data makes the analysis process more complex. We propose an unsupervised method for integrating multiple data views to produce a single unified graph representation, based on the combination of the $k$-nearest neighbour sets for users derived from each view. These views can be either relation-based or feature-based. The proposed method is evaluated on a number of annotated multi-view Twitter datasets, where it is shown to support the discovery of the underlying community structure in the data.
\end{abstract}

%-------------------------------------------------------------
%!TEX root = aggregation.tex
\section{Introduction}
\label{sec:intro}

Social networks are often represented using multiple \emph{views} or relations that share all, or part of the same user set. In many cases, these views will consist of graphs with heterogeneous edge types, where each type has different semantics, along with different frequency or weight distributions \cite{cai05hidden}. For instance, in the case of Twitter, we can characterise users by the accounts whom they follow (or who follow them), the users whom they retweet  (or who retweet them), the curated lists to which they have been assigned, and so on. In bibliographic networks, we can describe groups of researchers in terms of either their co-authorship relations or co-citation links.  Additionally, users in real-world social networks often have associated attribute information, such as demographic details or user-generated textual content (\eg the content of a user's tweets on Twitter; a collection of abstracts for papers published by a researcher). 

For many social network analysis tasks, it will be preferable to work with a unified representation that  summarises the information provided by all the data views, rather than working separately on the individual views. A wide variety of community finding algorithms have been proposed in the literature that assume the existence of a single relation between nodes \cite{fortunato10review}. However, increasingly there is interest in uncovering community structure from richer data sources that provide multiple relations \cite{tang12community}. 
From a visualisation perspective, it is much easier to interpret a graph with a single aggregated relation (as shown in Figs.~2-4), than it is to interpret representations that include multiple different types of relations (\eg retweet, follows, mentions). 
In the task of user curation on social media platforms, it is necessary to combine information from multiple views to produce a definitive set of recommendations \cite{greene12recsys}. 

In \refsec{sec:methods}, we propose a new method for integrating multiple data views to provide a sparse, unified graph representation, which retains the most informative connections from the original views. The aggregation process is performed at a local level, by combining the ranked neighbour sets for each individual user, and then constructing an overall directed nearest neighbour graph from the local neighbour sets. Unlike many alternative approaches, the views can be either relation-based or feature-based, once a similarity or ranking measure is defined on those views. The views can be incomplete (\ie not all users are present in each view), once there is a partial mapping between the views. Also, there is no requirement to manually select parameters indicating the relative importance of the different views, and no requirement for supervision in the form of labelled training examples.  In \refsec{sec:eval}, we describe evaluations on a collection of annotated Twitter datasets, which show that the unified graphs facilitate the identification of meaningful community structure from multi-view data.  In \refsec{sec:conc}, we conclude with a discussion of extensions and further applications for our proposed method.
%!TEX root = aggregation.tex
\section{Related Work}
\label{sec:related}

A range of techniques have been described for clustering across multiple feature-based views.
Bickel \& Scheffer \cite{bickel04multiview} introduced multi-view versions of the traditional $k$-Means and EM clustering algorithms, which operate by interleaving the optimisation processes for the different views.
Zhou \& Burges \cite{zhou07spectral} proposed a spectral clustering approach for application to multiple graphs sharing the same set of nodes, based on a mixture of Markov chains defined on the different views. The relative importance of each graph is defined by a manually-specified parameter. 
Greene \& Cunningham \cite{greene09ecir} proposed a ``late integration'' strategy for clustering heterogeneous data sources, based on the concept of cumulative voting in unsupervised ensembles. The strategy was applied to bibliographic data, consisting of co-citation relations and paper abstracts represented using a bag-of-words model.
More recently, Liu \etal \cite{liu13multi} proposed a joint non-negative matrix factorisation algorithm, which applies an iterative update procedure to find a consensus between the input matrices. The influence of each view on the outcome is determined by a user-specified set of regularisation parameters.

In the context of network analysis, the direct integration of multiple relation types can prove difficult, if the relations in the different views are not comparable, or if the relations in one view are considerably more sparse than another \cite{tang12community}. 
Cai \etal \cite{cai05hidden} emphasised the importance of mining heterogeneous relations in social networks to identify hidden groups. The authors proposed a regression-based technique to find the optimal linear combination of a number of different weighted relation matrices, relying on a set of input examples that have been assigned community labels. Based on the combined relations, the authors then applied a spectral clustering algorithm to produce disjoint communities.
Recently, Gollini \& Murphy \cite{gollini13arxiv} proposed an extension of existing latent space models for jointly modelling information from multiple network link relations on a given set of nodes. To fit the model, the authors use a variational Bayes inference approach, supporting the analysis of up to thousands of nodes.  

While most community finding algorithms assume the existence of only one kind of relation, Tang~\etal~\cite{tang12community} focused on the problem of finding groups of related users in ``multi-dimensional networks''. The authors described a range of alternative strategies, including modularity-based community finding applied to the average interaction network among a group of users, a ``feature integration'' strategy where structural features from different views are mapped  into the same compatible space, and a strategy based on an ensemble of clusterings generated on different views. These alternatives were evaluated on synthetic data and a dataset of YouTube users represented via five network views, where the feature integration strategy was shown to be most effective.
%!TEX root = aggregation.tex
\section{Methods}
\label{sec:methods}

We now propose a method to produce a unified network representation from either feature-based or relational views on a set of social network users. Specifically, we propose the application of SVD rank aggregation to a matrix encoding multiple nearest neighbour sets for each user. This form of rank aggregation has been previously used in identifying anomalous behaviour~\cite{wu10recsys}, and for recommending users in list curation~\cite{greene12recsys}.  The resulting aggregated per-user rankings are then combined to form a global graph covering all users. This sparse graph represents a unified summarisation of the strongest connections between users across all views.

\subsection{Neighbour Set Identification}
\label{sec:methods1}
\begin{figure*}[!t]
\centering
\subfigure[\emph{view 1}]{
\includegraphics[width=0.175\linewidth]{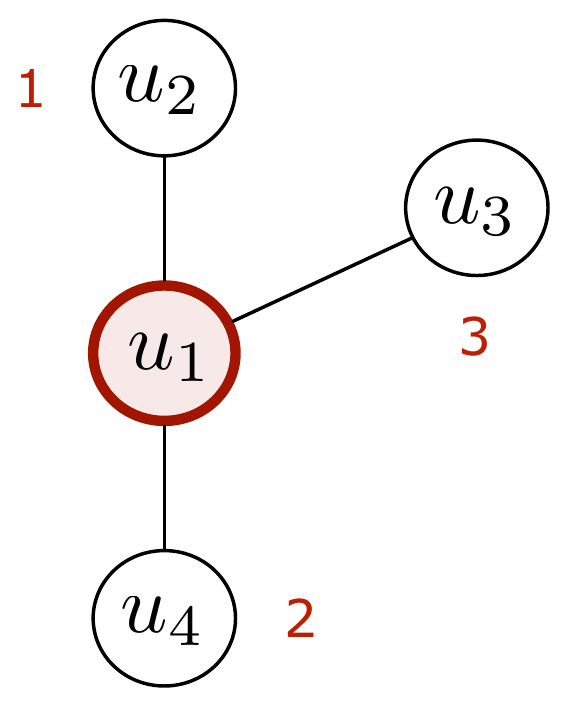}
}
\hskip 1.15em
\subfigure[\emph{view 2}]{
\includegraphics[width=0.175\linewidth]{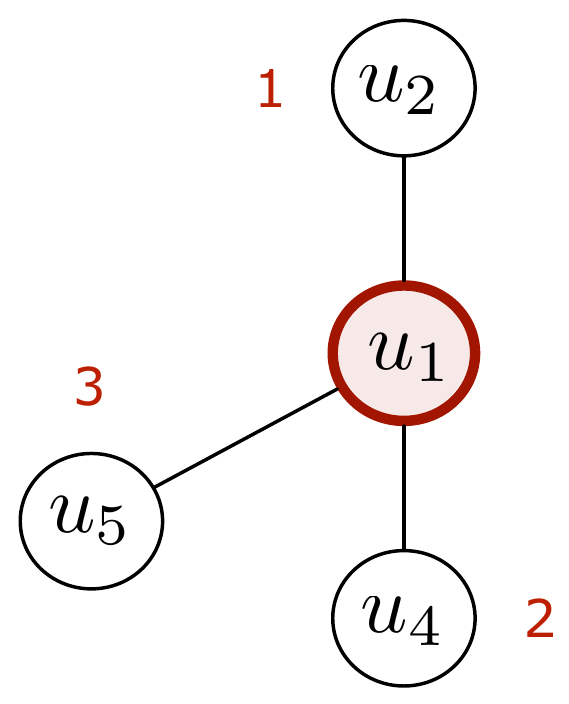}
}
\hskip 2.1em
\subfigure[\emph{view 3}]{
\includegraphics[width=0.173\linewidth]{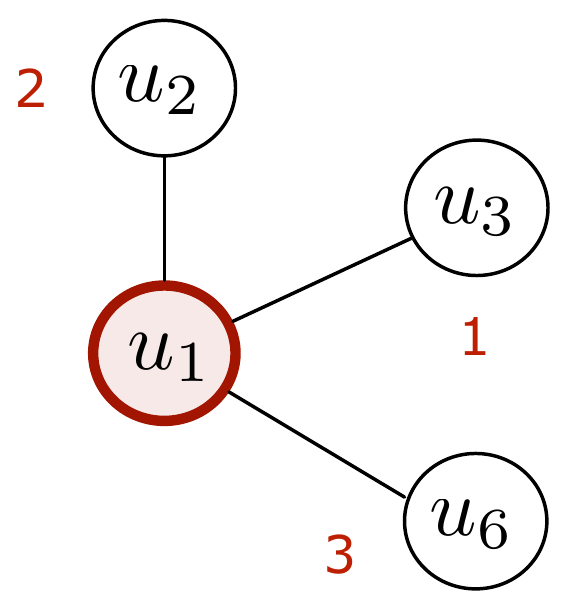}
}
\hskip 1.54em
\subfigure[\emph{aggregated}]{
\includegraphics[width=0.253\linewidth]{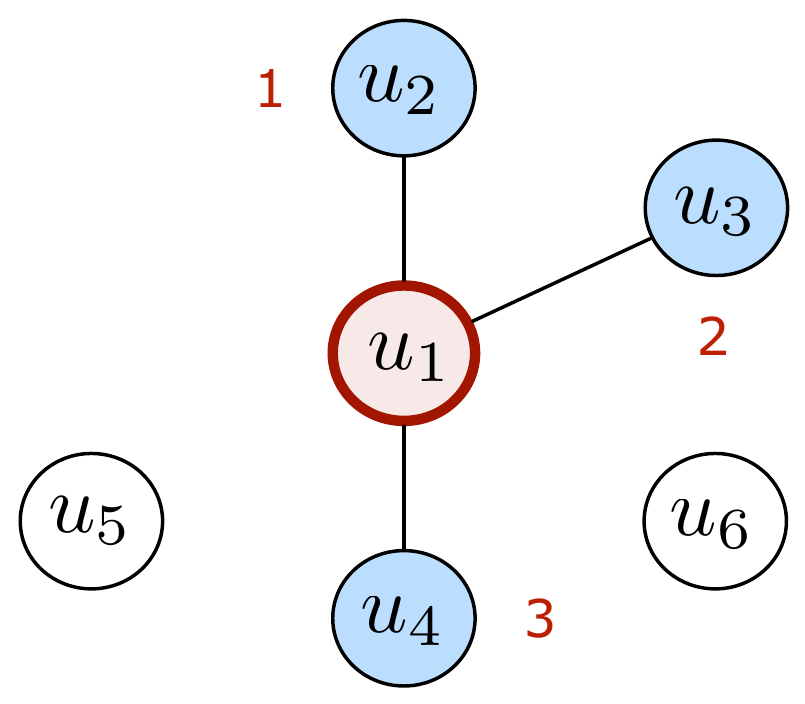}
}
\caption{Example of the proposed aggregation method, involving six users and three views.
Graphs (a)-(c) show the ranked neighbour sets for the user $u_{1}$ for $k=3$. By combining the ranks from these neighbourhoods, we produce the aggregated neighbour set $\{u_{2},u_{3},u_{4}\}$ for $u_{1}$, as shown in (d).}
\label{fig:example}
\end{figure*}

The input to the aggregation process is a dataset of users $\fullset{u}{n}$, along with $l$ different views, each representing some or all of the $n$ users. These views may be relation-based or feature-based. The only requirement is that some measure of similarity is provided for each view -- either a metric or non-metric measure can be used. The only parameter required for the aggregation process is a value for the number of nearest neighbours $k$. This value controls the sparsity of the output graph -- a lower value of $k$ will result in a less dense graph.
The first phase of the aggregation process is as follows, for each user $u_{i}$:
\begin{enumerate}
\item For each view $j = 1\;to\;l$, compute a similarity vector $\vec{v}_{ij}$ between $u_{i}$ and all other users present in that view, using the similarity measure provided for the view.
\item From the values in $\vec{v}_{ij}$, produce a rank vector of all other $(n-1)$ users relative to $u_{i}$, denoted $\vec{r}_{ij}$. In cases where not all users are present in view $j$, missing users are assigned a rank of $(n_{j}'+1)$, where $n_{j}'$ is the number of users present in the view. 
\item Stack all $l$ rank vectors as columns, to form the $(n-1) \times l$ rank matrix $\m{R}_{i}$, and normalise the columns of this matrix to unit length.
\item Compute the SVD of $\mtr{R}_{i}$, and extract the first left singular vector. Arrange the entries in this vector in descending order, to produce a ranking of all other $(n-1)$ users.
Then select the $k$ highest ranked users as the \emph{neighbour set} of $u_{i}$.
\end{enumerate}
A simple example illustrating the method is shown in \reffig{fig:example}. The procedure can be readily parallelised by processing multiple users simultaneously. In addition, the  time required for the  aggregation process can be reduced considerably by computing the truncated SVD of the rank matrices. While many alternative rank aggregation techniques exist that could be used in conjunction with our method (\eg \cite{klementiev07ranking}), we choose  SVD ranking for computational reasons. 

\subsection{Unified Graph Construction}
\label{sec:methods2}
Once the $k$-nearest neighbour sets have been identified  for all $n$ users, we use this information to build a global graph representation of the dataset. A natural approach to combine the sets is to construct the corresponding asymmetric $k$-\emph{nearest-neighbour graph}.  Specifically, we construct a directed unweighted graph, where each node is a user and an edge exists from node $i$ to $j$, if $u_{j}$ is contained in the neighbour set of $u_{i}$. This process yields a sparse, unified graph encoding the connectivity information derived from all original views in the dataset, representing all users that were present in one or more of those views.

Note that, since the neighbour relations produced using the above method are not necessarily symmetric, it can be the case that an edge exists from $u_{i}$ to $u_{j}$, but not from $u_{j}$ to $u_{i}$. In applications where directed graphs are not appropriate (\eg for certain community finding algorithms), we can readily build an undirected mutual nearest neighbour graph based on shared neighbours \cite{gowda78agglomerative}.

%!TEX root = aggregation.tex
\section{Evaluation}
\label{sec:eval}

We now examine the degree to which the aggregation method proposed in \refsec{sec:methods} preserves the most informative underlying associations between users in the original views, as determined by ground truth community assignments.

\subsection{Datasets}
\label{sec:data}

For our evaluation, we collected five new Twitter datasets, for which sets of manually-curated ground truth communities are available.
\begin{description}
\item[\emph{football}:] A collection of 248 English Premier League football players and clubs active on Twitter. The disjoint ground truth communities correspond to the 20 individual clubs in the league.
\item[\emph{olympics}:] A dataset of 464 users\footnote{A subset of a Twitter user list originally curated by The Telegraph in 2012:\\\url{https://twitter.com/Telegraph2012/london2012}}, covering athletes and organisations that were involved in the London 2012 Summer Olympics. The disjoint ground truth communities correspond to 28 different sports.
\item[\emph{politics-ie}:] A collection of Irish politicians and political organisations, assigned to seven disjoint ground truth groups, according to their affiliation.
\item[\emph{politics-uk}:] 419 Members of Parliament (MPs) in the United Kingdom. The ground truth consists of five groups, corresponding to political parties.
\item[\emph{rugby}:] A collection of 854 international Rugby Union players, clubs, and organisations currently active on Twitter. The ground truth consists of overlapping communities corresponding to 15 countries. In the case of players, these user accounts can potentially be assigned to both their home nation and the nation in which they play club rugby.
\end{description}
Summary statistics for the five datasets are provided in \reftab{tab:data}. Pre-processed versions of these datasets are available online\footnote{See \url{http://mlg.ucd.ie/networks}}.

\begin{table*}[!h]
\centering
\caption{Summary of Twitter datasets used in our evaluations, including total number of users, tweets, user lists, and the number of associated ground truth communities.}
\begin{tabular}{|l|c|c|c|c|}\hline
\bf \;Dataset & \bf \;\;\;\# Users\;\; & \bf \;\;\# Tweets\;\; & \bf \;\# User Lists\;& \bf \# Communities \\ \hline
\it \;football & 248 & 351,300 & 7,814 & 20 \\ 
\it \;olympics & 464 & 725,662  & 4,942 & 28 \\ 
\it \;politics-ie\;\;\; & 348 & 267,488 & 1,047 & 7 \\ 
\it \;politics-uk\;\;\;\;\;\; & 419 & 539,592 & 3,614 & 5 \\ 
\it \;rugby & 854 & 1,166,379 & 5,900 & 15 \\ \hline
\end{tabular}
\label{tab:data}
\end{table*}

For each dataset, we constructed a heterogeneous collection of nine network- and content-based views, containing some or all of the complete set of Twitter users for that dataset. In all cases, cosine similarity is applied to the representation to produce the pairwise similarities used in the aggregation process. For a more detailed explanation of the representations listed below, consult \cite{greene12recsys}.
\begin{description}
\item[1. tweet content:] User content profiles, constructed from the concatenation of the 500 most recently-posted tweets for each user.
\item[2. list text:] List content profiles, constructed from the concatenation of both the \emph{names} and the \emph{descriptions} of the 500 Twitter lists to which each user has most recently been assigned.
\item[3. follows:] From the unweighted directed follower graph, construct binary user profile vectors based on the users whom they follow (\ie out-going links).
\item[4. followed-by:] From the unweighted directed follower graph, construct binary user profile vectors based on the users that follow them (\ie incoming links). A pair of users are deemed to be similar if they are frequently ``co-followed'' by the same users. 
\item[5. mentions:] From the weighted directed mention graph, construct user profile vectors based on the users whom they mention.
\item[6. mentioned-by:] From the weighted directed mention graph, construct binary user profile vectors based on the users that mention them.  A pair of users are deemed to be similar if they are frequently ``co-mentioned'' by the same users. 
\item[7. retweets:] From the weighted directed retweet graph, construct user profile vectors based on the users whom they retweet.
\item[8. retweeted-by:] From the weighted directed retweet graph, construct user profile vectors based on the users that retweet them.  Users are deemed to be similar if they are frequently ``co-retweeted'' by the same users. 
\item[9. co-listed:] Based on Twitter user list memberships, construct an unweighted bipartite graph, such that an edge between a list and a user indicates that the list contains the specified user.  A pair of users are deemed to be similar if they are frequently linked to the same lists. Again, we only consider the 500  lists to which each user has been assigned most recently assigned.
\end{description}

\subsection{Evaluation Measures}
\label{sec:measure}

Ding \& He \cite{ding04knn} formalised the concept of \emph{$k$-nearest-neighbour consistency} for clustering. That is, for any item in a cluster, its $k$-nearest neighbours should also be assigned to the same cluster. Motivated by this work, given a ground truth set of user communities, we evaluate the degree to which alternative views preserve the $k$-nearest-neighbour consistency of those communities. 
A representation of the data that corresponds well to the ground truth will have a high level of consistency, while a representation that does not preserve the structure of the ground truth will yield a low level of consistency. 

For a single user $u_{i}$ and view, we can compute the \emph{user consistency} as the fraction of that user's $k$ nearest neighbours in that view that are assigned to the same ground truth community. In the case of overlapping ground truth communities, we generalise by counting the fraction of neighbours that are assigned to at least one community also containing $u_{i}$.
For a complete dataset, we compute the \emph{micro-average consistency} as the simple average of all $n$ user consistency scores. 
As with traditional micro-averaged classification accuracy, this will tend to reflect performance on larger communities, where community sizes are unbalanced. Therefore, we also compute the \emph{macro-average consistency} as follows: for each ground truth community, we calculate the average user consistency for users assigned to that community; an overall score for the dataset is calculated as the simple average of all community scores. Thus, in the macro measure, both small and larger ground truth communities are weighted equally.

\begin{figure*}[!b]
\centering
\includegraphics[width=0.75\linewidth]{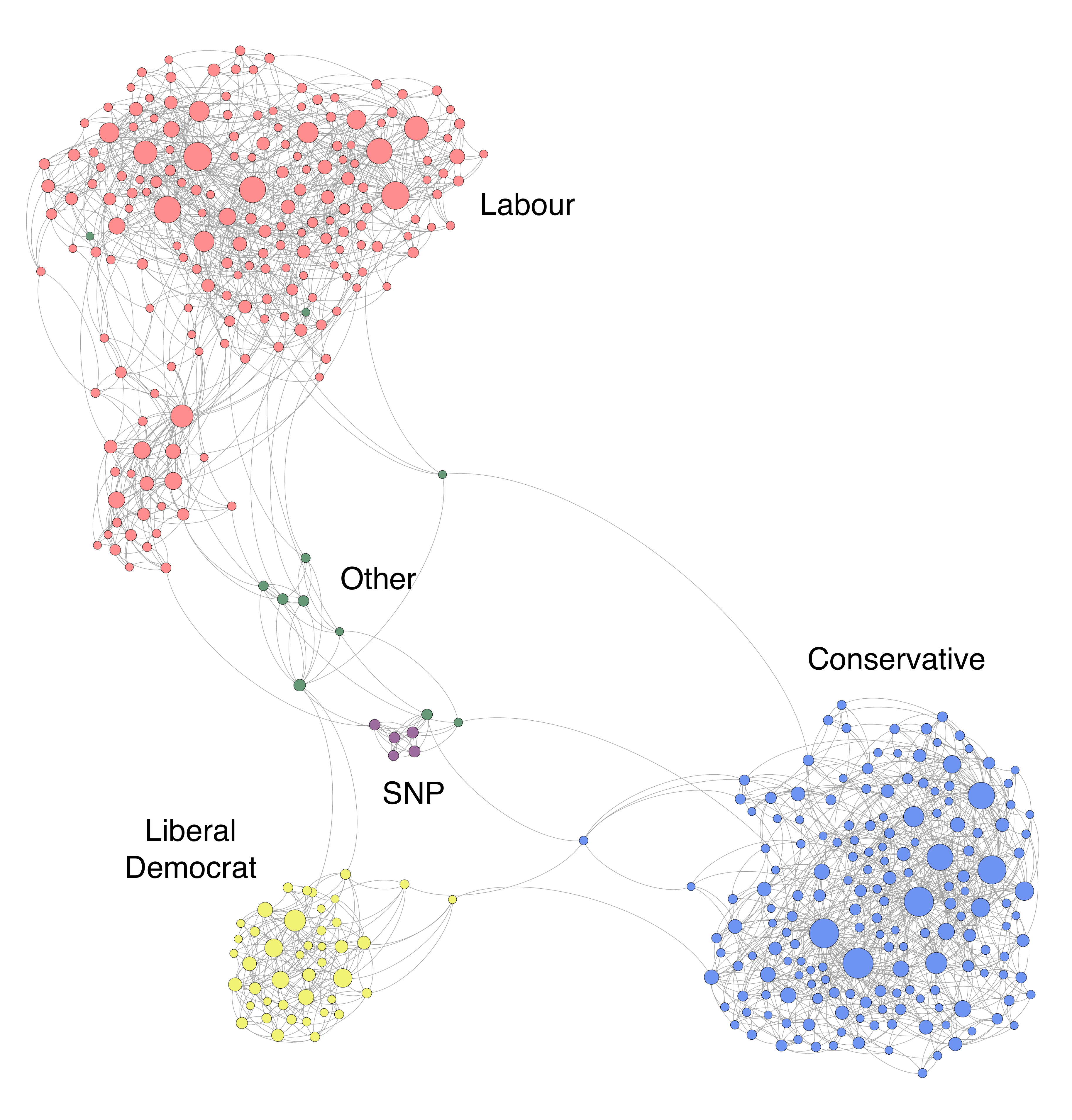}
\vskip -1em
\caption{Unified graph constructed from nine views of the \emph{politics-uk} dataset $(k=5)$. Users are coloured and labelled based on a ground truth, corresponding to five different political groupings. }
\label{fig:ukmps}
\end{figure*}

\begin{figure*}[!b]
\centering
\includegraphics[width=0.93\linewidth]{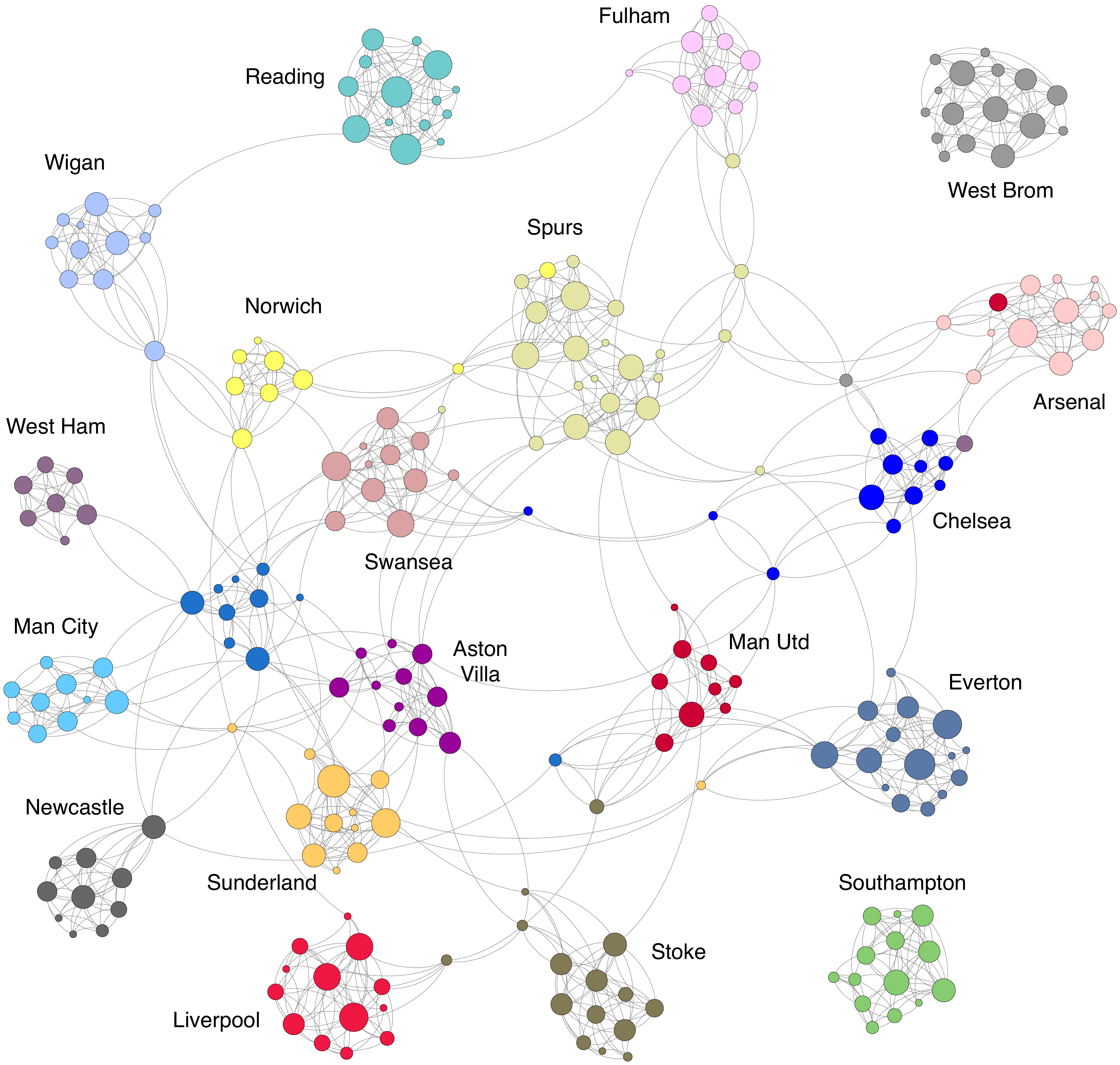}
\vskip -0.5em
\caption{Unified graph constructed from nine views of the \emph{football} dataset $(k=5)$. Groups of users are coloured and labelled based on a ground truth, corresponding to 20 English Premier League clubs.}
\label{fig:football}
\end{figure*}

\begin{figure*}[!b]
\centering
\includegraphics[width=0.89\linewidth]{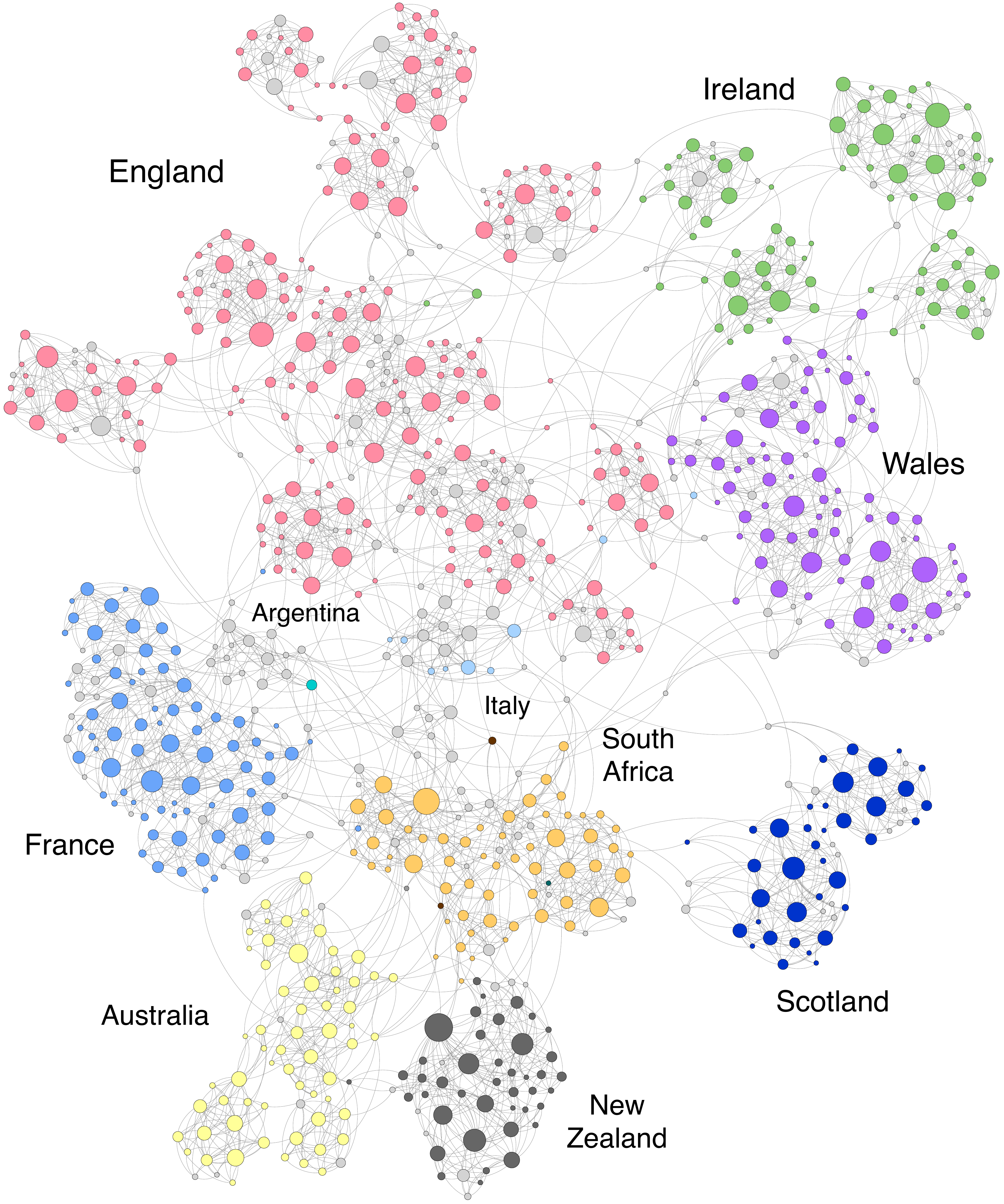}
\vskip -0.5em
\caption{Unified graph constructed from nine views of the \emph{rugby} dataset $(k=5)$. Users are coloured and labelled based on a ground truth, corresponding to 15 different countries. Users assigned to multiple communities are coloured in grey.}
\label{fig:rugby}
\end{figure*}

\subsection{Discussion}
\label{sec:discuss}

When we apply our proposed approach to the five Twitter datasets, a visual inspection of the output (based on force-directed layouts produced using Gephi \cite{bastian09gephi}) highlights the sparsity of the unified graphs. In a number of cases we see almost entirely disconnected components, where users assigned to the same ground truth communities are densely-connected, while there is little connectivity between those communities. This is particularly clear in the case of the two political datasets. For instance, in \reffig{fig:ukmps} we see that, for the unified graph at $k=5$, there is a clear separation between the various political groupings, with only a handful of long-range inter-community links in the graph. Given that we can see the separation clearly by visual inspection alone, any reasonable single-mode community finding algorithm should be able to identify this grouping.

It is interesting to note that our approach also supports the discovery of sub-communities relative to the ground truth, which had not been identified manually. In \reffig{fig:ukmps}, we observe that the community for the Labour Party contains a smaller sub-community of users. On inspecting these accounts, it is apparent that they correspond to Labour Party MPs based in Scotland. \reffig{fig:football} shows that the aggregated approach clearly divides the data in the \emph{football} dataset according to the different Premier League clubs. The small number of weak ties between clubs almost all indicate players who have recently been transferred or loaned between clubs, leading to some conflicting information, particularly in the cases of their less recent tweet content and user list assignments. 

\begin{figure*}[!b]
\centering
\subfigure[\emph{football}]{
\includegraphics[width=0.468\linewidth]{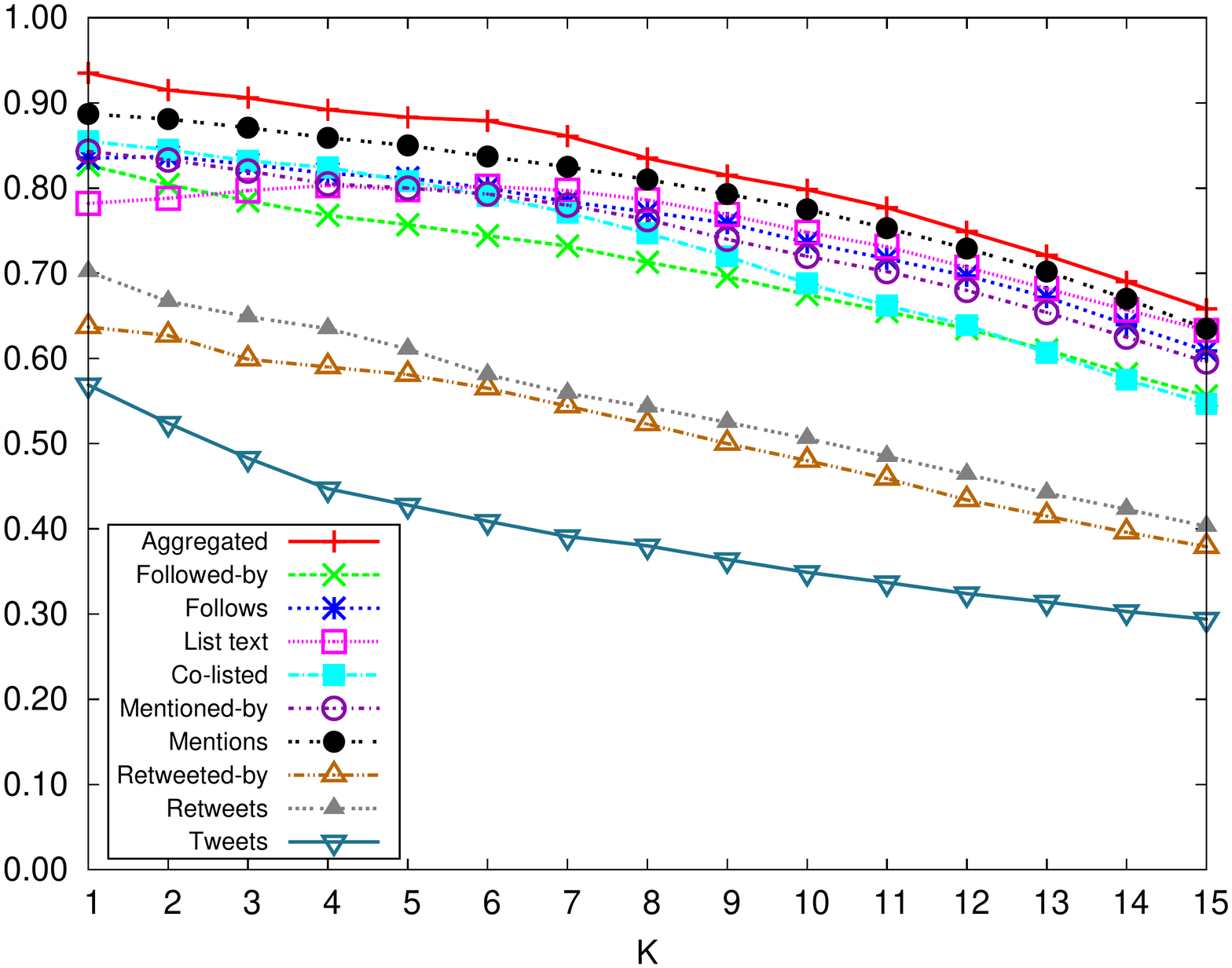}
}
\subfigure[\emph{olympics}]{
\includegraphics[width=0.468\linewidth]{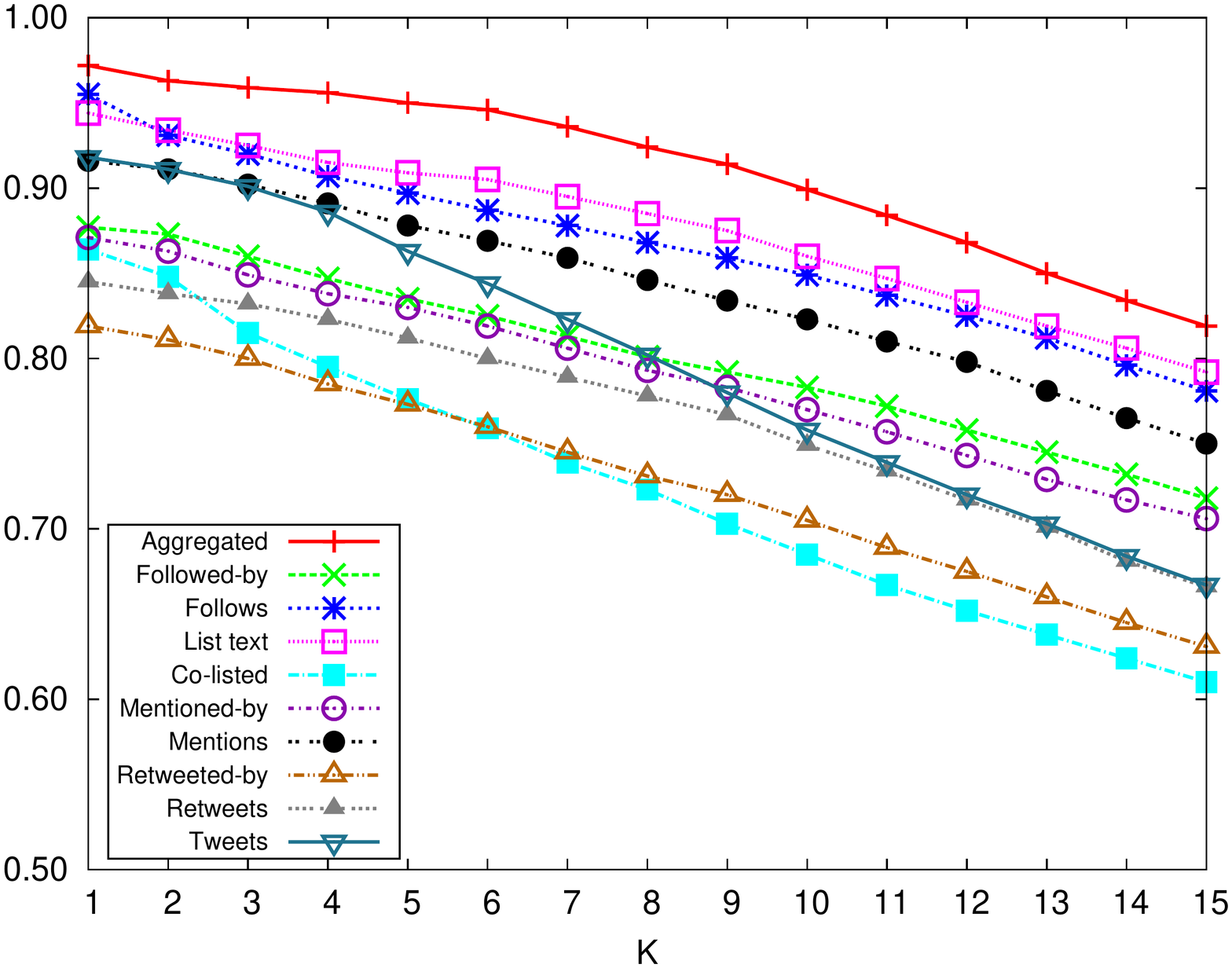}
}
\subfigure[\emph{politics-ie}]{
\includegraphics[width=0.468\linewidth]{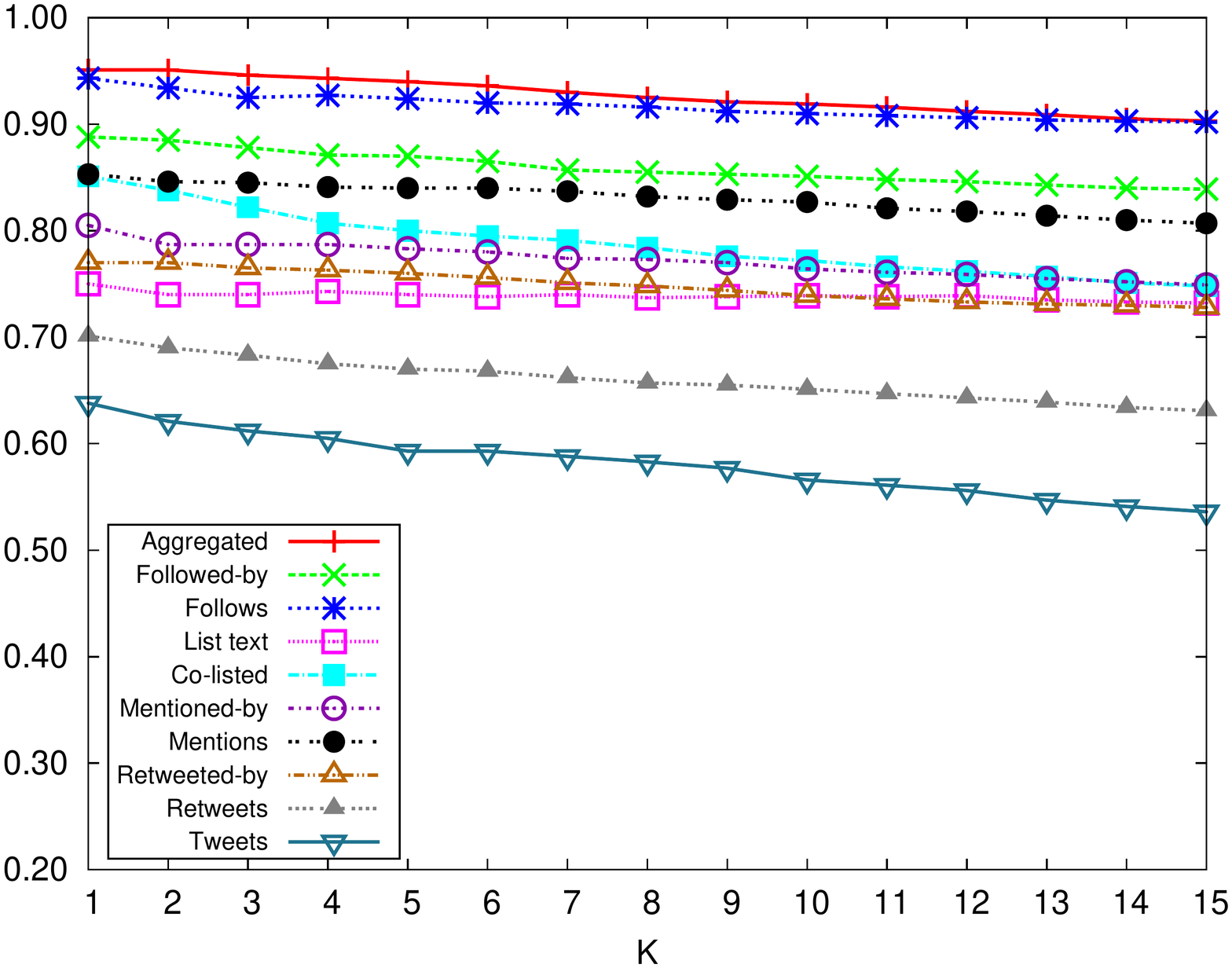}
}
\subfigure[\emph{politics-uk}]{
\includegraphics[width=0.468\linewidth]{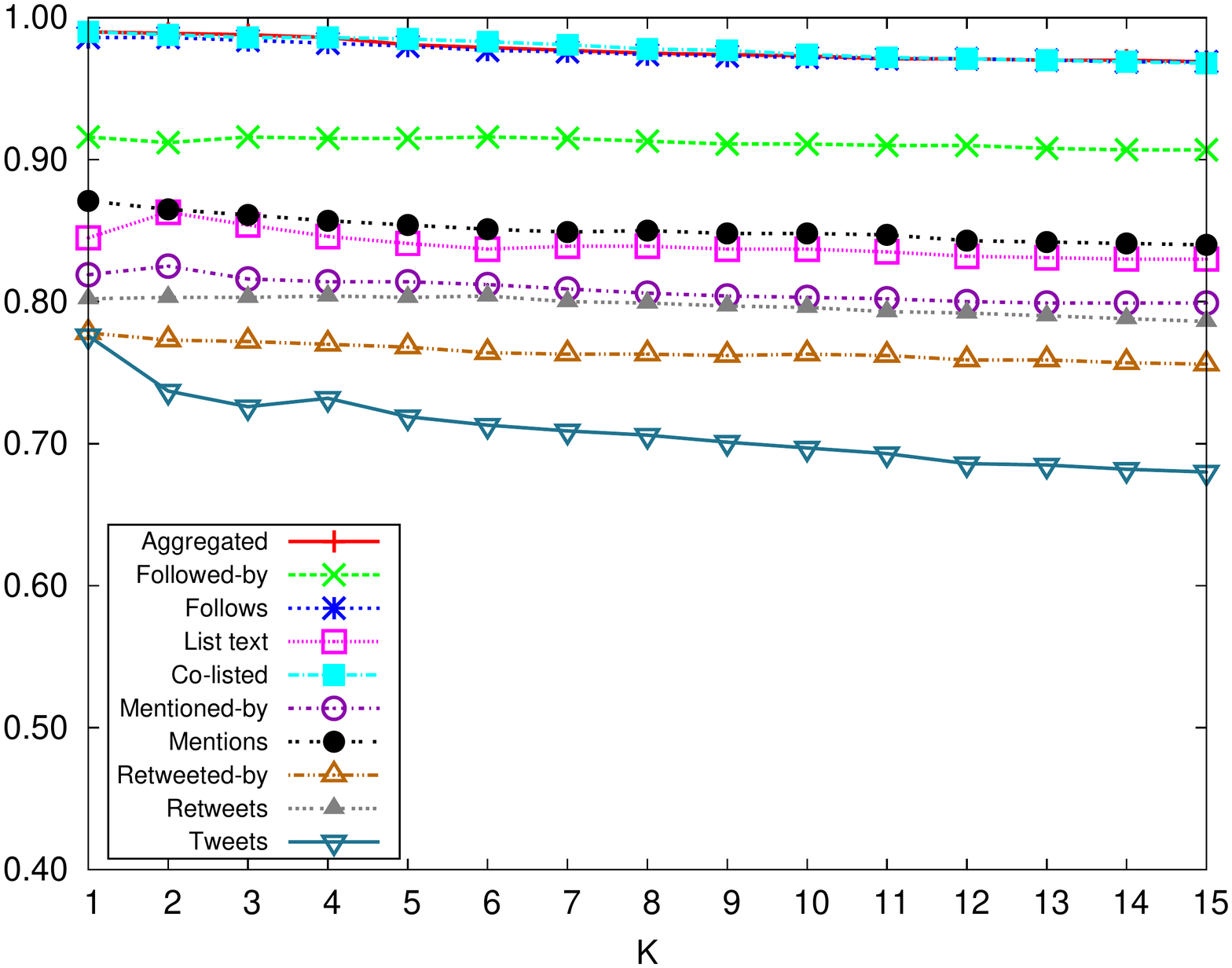}
}
\subfigure[\emph{rugby}]{
\includegraphics[width=0.468\linewidth]{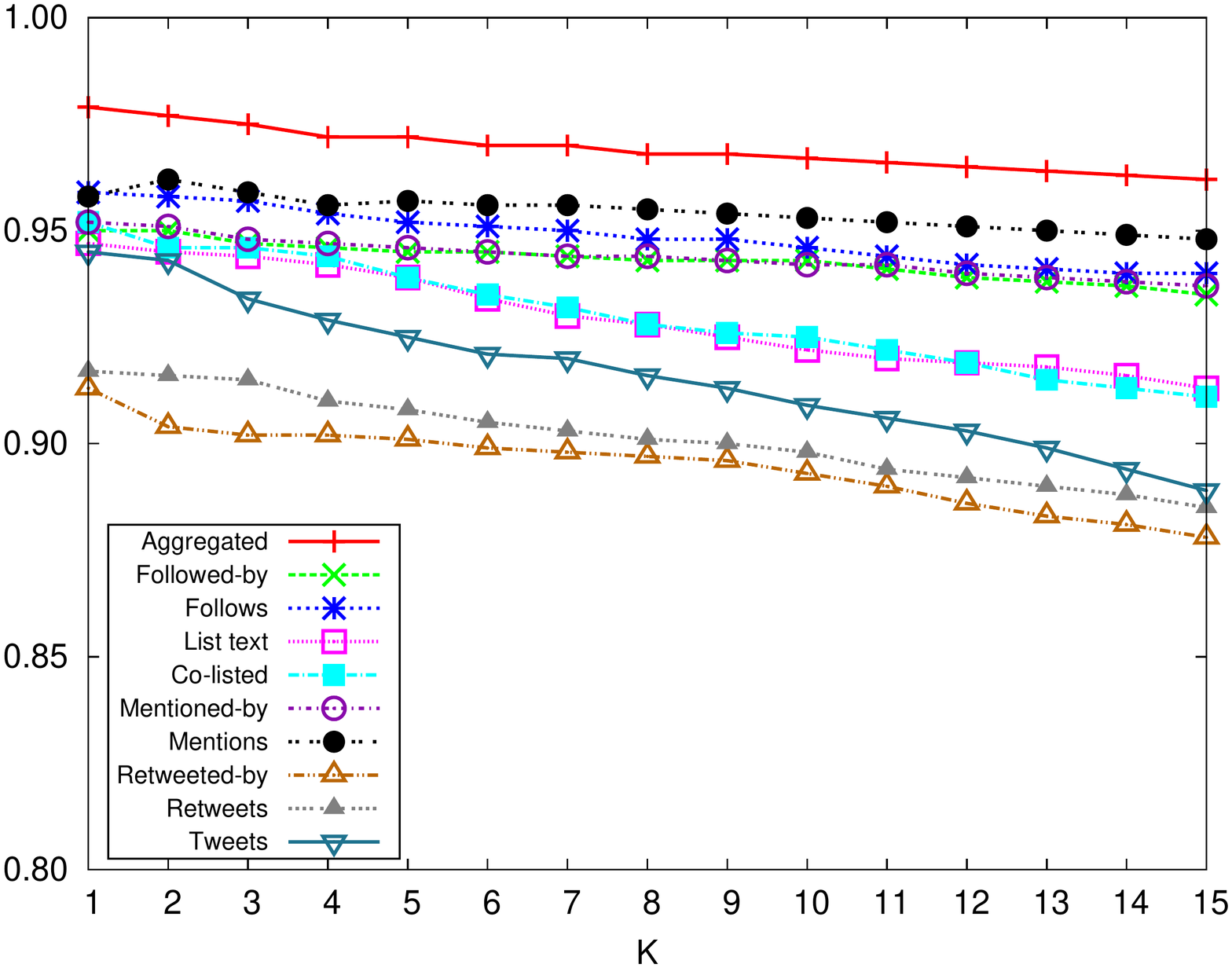}
}
\vskip -0.5em
\caption{Comparison of \emph{micro-average consistency} scores for $k \in [2,15]$, calculated on nine individual views and the resulting unified graph, across five Twitter datasets.}
\label{fig:micro}
\end{figure*}

A more complex picture is visible in the case of the \emph{rugby} dataset in \reffig{fig:rugby}. The presence of overlapping nodes generally leads to a higher level of connectivity between communities. Again we see strong evidence for sub-communities within the ground truth communities -- in particular we see sub-groups corresponding to individual clubs based in England, Ireland, Scotland, and Australia. Also, we see that certain nodes lying between communities correspond to players who have recently transferred between clubs located in different countries. Overall, we observed similar patterns of clustering behaviour across all the datasets, for a wide range of $k$ values.

\begin{figure*}[!b]
\centering
\subfigure[\emph{football}]{
\includegraphics[width=0.468\linewidth]{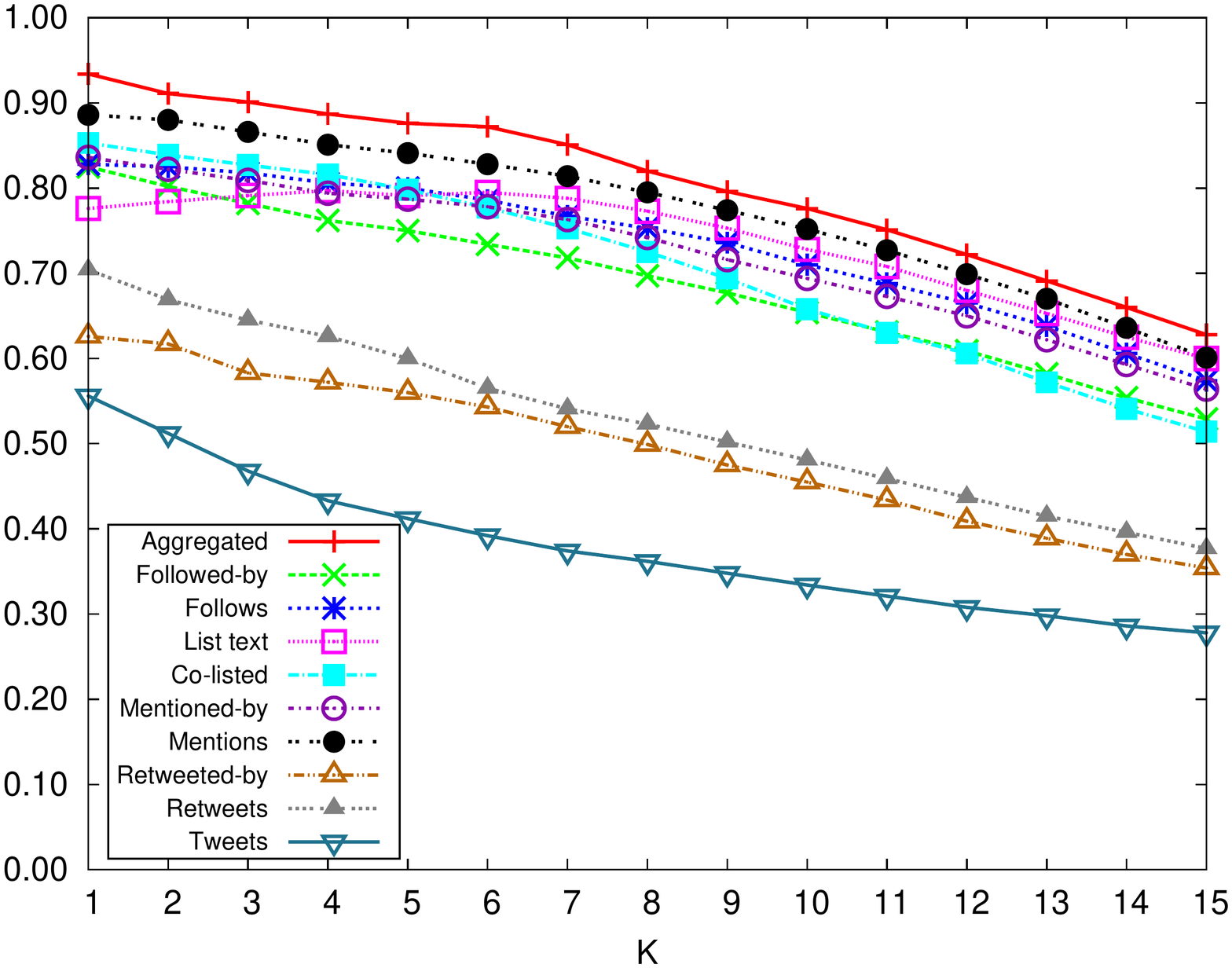}
}
\subfigure[\emph{olympics}]{
\includegraphics[width=0.468\linewidth]{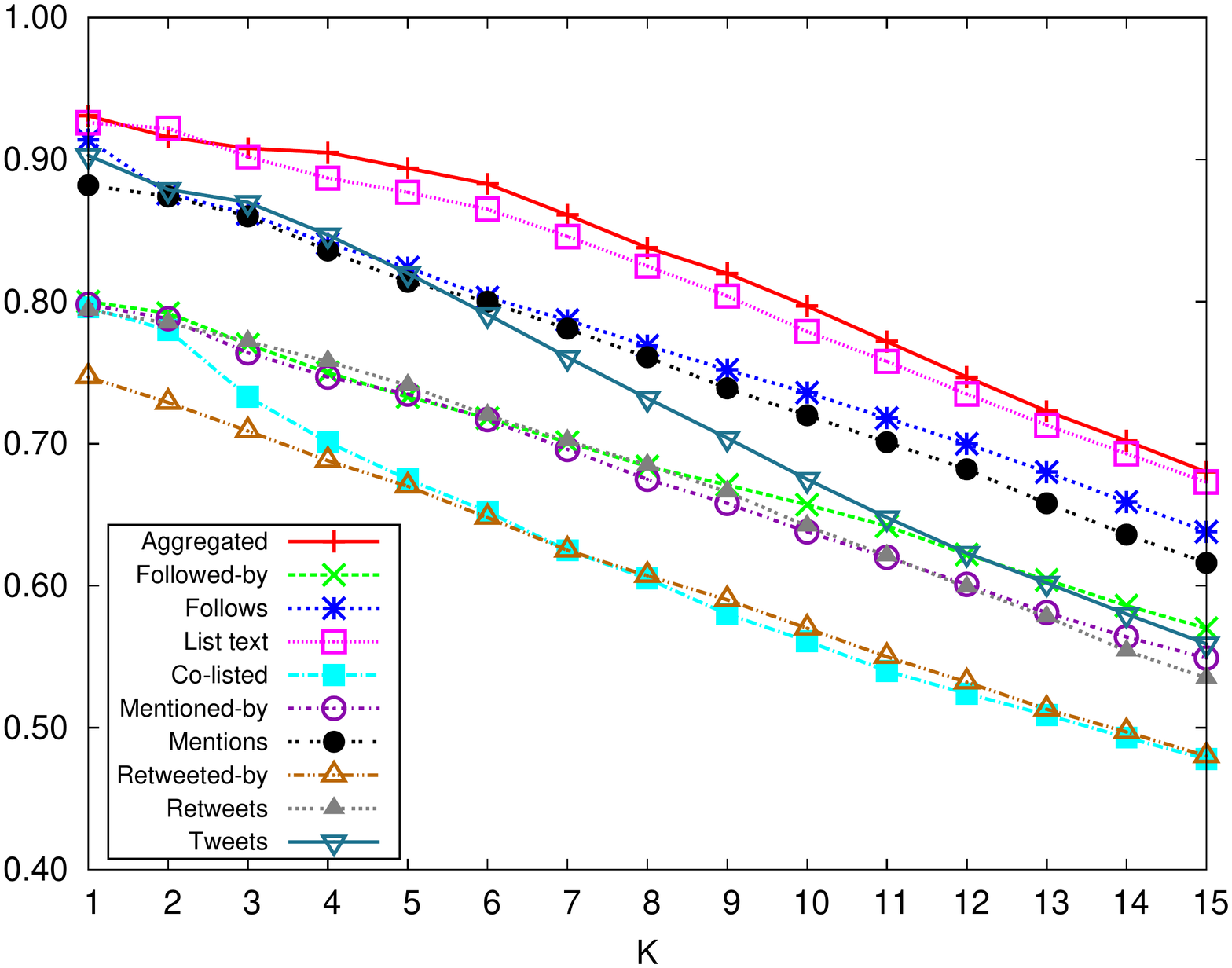}
}
\subfigure[\emph{politics-ie}]{
\includegraphics[width=0.468\linewidth]{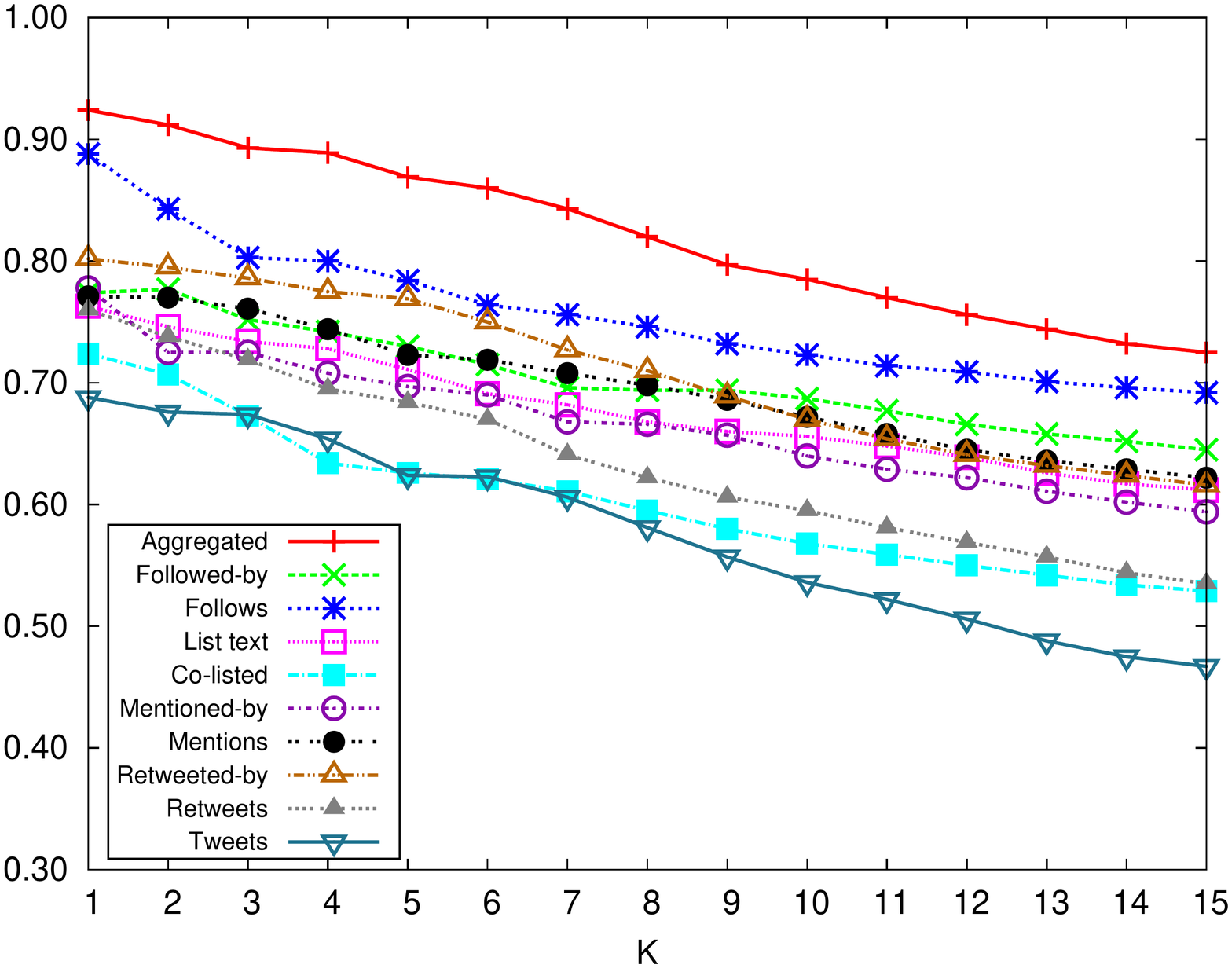}
}
\subfigure[\emph{politics-uk}]{
\includegraphics[width=0.468\linewidth]{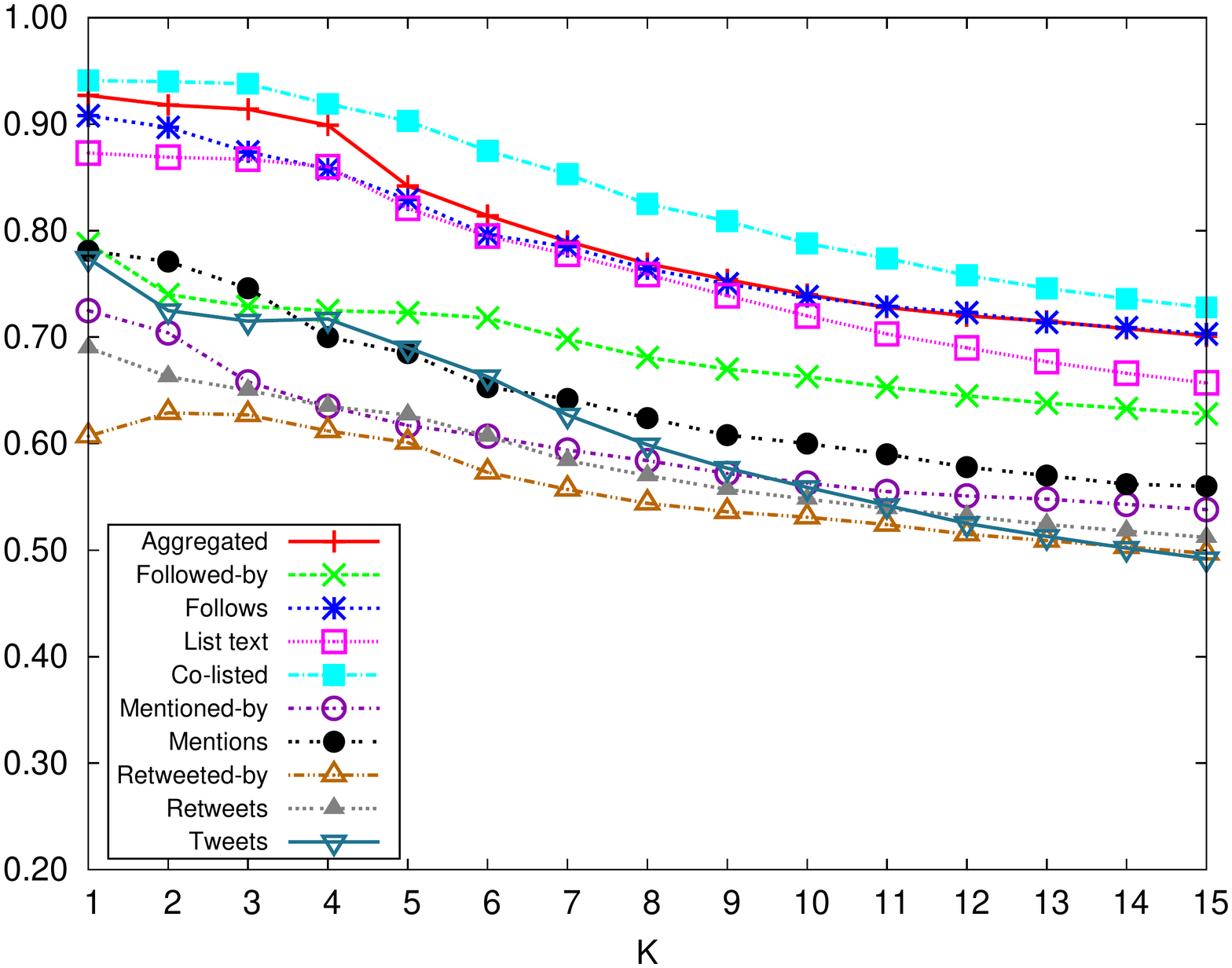}
}
\subfigure[\emph{rugby}]{
\includegraphics[width=0.468\linewidth]{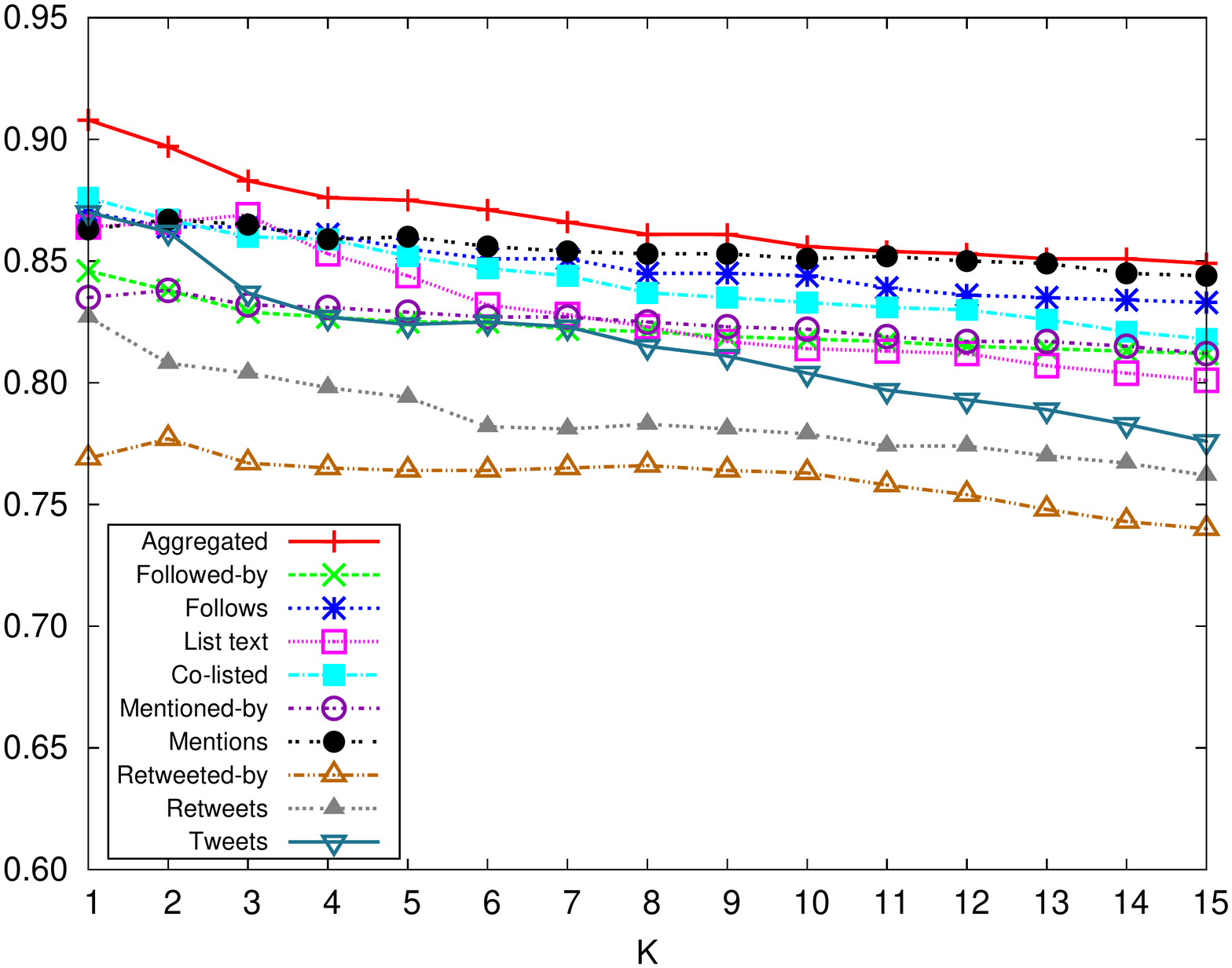}
}
\vskip -0.5em
\caption{Comparison of \emph{macro-average consistency} scores for $k \in [2,15]$, calculated on nine individual views and the resulting unified graph, across five Twitter datasets.}
\label{fig:macro}
\end{figure*}

To quantitatively analyse the effectiveness of our method, we calculated micro-average and macro-average consistency scores, for neighbourhood sizes $k \in [2,15]$. We then compared the scores afforded by the $k$-nearest neighbours for the individual views with those achieved by the unified graph. 
\reffig{fig:micro} shows the micro-average consistency scores for the nine individual views, plus the aggregated representation. In four of the five datasets, the unified graph provides a higher level of consistency among neighbours than any of the individual views. It is only in the case of the \emph{politics-uk} dataset that the co-listed view out-performs the aggregated approach. Here it appears that there is a high number of carefully-curated user lists on Twitter corresponding to UK political party memberships, while the aggregation method is somewhat affected by the poor quality of information provided by tweet content. 
In \reffig{fig:macro}, the macro-average consistency scores for experiments on the five datasets are shown. We see similar trends for the datasets as in \reffig{fig:micro}, with the aggregation method performing well, even in the presence of unbalanced ground truth community sizes, such as in the \emph{politics-ie} and \emph{olympics} datasets. Again, it is only in the case of the \emph{politics-uk} dataset where an individual view achieves higher scores.

One key observation that can be made from the above results is that no single individual view out-performs all others on every dataset. So while, for example, user lists are highly-informative in the case of the \emph{politics-uk} dataset, user list information proves far less useful for identifying distinct political groupings in the case of the \emph{politics-ie} dataset. In general, we will typically not know \apriori which individual view is most informative for a given dataset. This will prove problematic for multi-view analysis methods that require the relative importance of each view to be specified as an input parameter. In contrast, our proposed method does not require the manual prioritisation or weighting of the constituent views, and performed robustly across all datasets in our experiments. 
 
%!TEX root = aggregation.tex
\section{Conclusions}
\label{sec:conc}
We have demonstrated that we can use a form of rank aggregation applied to nearest neighbour sets to construct a single unified graph from multiple heterogeneous data views. Evaluations on a number of annotated Twitter datasets have shown that the unified graphs can preserve and highlight the underlying community structure in the data. We suggest that this procedure will prove useful as a pre-processing step prior to other network analysis tasks, such as community finding, visualisation, and user recommendation. 

In our current model, each user in the aggregated graph has at most $k$ outgoing edges. In future work, we plan to examine the adaptive selection of $k$ on a per-user basis, to allow for hub users with many connections, or outlying users with little connectivity in any of the data views. We also plan to expand our evaluations to include the use of unified graphs in conjunction with existing community finding algorithms. Our method also has potential applications in cross-network analysis, to support the combination of partial views from multiple social media platforms, such as Twitter and Facebook.

%-------------------------------------------------------------

\vspace{3 mm}\noindent\emph{Acknowledgments.}\;This research was supported by Science Foundation Ireland Grant 08/SRC/I1407 (Clique: Graph and Network Analysis Cluster).

\bibliographystyle{abbrv}
\bibliography{aggregation} 

\end{document}